\documentclass[10pt]{bmc_article}    

\usepackage{url}  
\usepackage{ifthen}  
\usepackage{multicol}   
\usepackage[utf8]{inputenc} 
\usepackage{amssymb}
\usepackage{graphicx}
\usepackage{amssymb}
\usepackage[cmex 10]{amsmath}
\usepackage{mdwmath}
\usepackage{multirow}
\usepackage{setspace}
\usepackage{needspace}
\usepackage{color}
\usepackage{amsthm}
\usepackage{tabu}
\usepackage{array}

\newcommand{\beq}{\begin{equation}}
\newcommand{\eeq}{\end{equation}}
\newcommand{\bqa}{\begin{eqnarray}}
\newcommand{\eqa}{\end{eqnarray}}

\newboolean{publ}

\begin{document}

\title{Improved mirror position estimation using resonant quantum smoothing}
 
\author{
 Trevor A. Wheatley\correspondingauthor$^{1,2}$
  \email{Trevor A Wheatley - t.wheatley@adfa.edu.au}
\and
  Mankei Tsang$^{3,4}$%
  \email{Mankei Tsang - mankei@nus.edu.sg}
\and
   Ian R. Petersen$^1$
  \email{Ian R Petersen - i.petersen@adfa.edu.au}
\and
  Elanor H. Huntington$^{1,2,5}$
  \email{Elanor H Huntington - elanor.huntington@anu.edu.au}
}
      
\address{
   \iid(1)School of Engineering and Information Technology, UNSW Australia, Canberra, ACT 2600, Australia.\\
    \iid(2)Centre for Quantum Computation \& Communication Technology, Australian Research Council, Canberra, Australia.\\
    \iid(3)Department of Electrical and Computer Engineering, National University of Singapore, 4 Engineering Drive 3, Singapore 117583, Singapore. \\
    \iid(4)Department of Physics, National University of Singapore, 2 Science Drive 3, Singapore 117551, Singapore. \\
    \iid(5)Research School of Engineering, College of Engineering and Computer Science, Australian National University, Canberra, ACT 2601, Australia.
}

\maketitle


\begin{abstract}
Quantum parameter estimation, the ability to precisely obtain a classical value in a quantum system, is very important to many key quantum technologies. Many of these technologies rely on an optical probe, either coherent or squeezed states to make a precise measurement of a parameter ultimately limited by quantum mechanics. We use this technique to theoretically model, simulate and validate by experiment the measurement and precise estimation of the position of a cavity mirror. In non-resonant systems, the achieved estimation enhancement from quantum smoothing over optimal filtering has not exceeded a factor two, even when squeezed state probes were used. Using a coherent state probe, we show that using quantum smoothing on a mechanically resonant structure driven by a resonant forcing function can result significantly greater improvement in parameter estimation than with non-resonant systems. In this work, we show that it is possible to achieve a smoothing improvement by a factor in excess of three times over optimal filtering. By using intra-cavity light as the probe we obtain finer precision than has been achieved with the equivalent quantum resources in free-space.
\end{abstract}

\ifthenelse{\boolean{publ}}{\begin{multicols}{2}}{}
\section*{Background}
\subsection*{Introduction}
The field of quantum metrology can be described as using quantum resources to enhance measurement precision beyond that achievable with purely classical resources. There are a number of resources that are available such as entanglement \cite{PRA_bollinger1996}, superposition \cite{Pryde}  and squeezing \cite{Nat-Breit1997}. There are also tools such as adaptive feedback \cite{wiseman_PRL_75} and quantum smoothing \cite{Tsang_smooth} to further exploit the quantum enhancement. Quantum parameter estimation (QPE) is a related discipline that is focussed more specifically on precisely estimating the classical parameters of a quantum system. The importance of QPE to fields such as gravitational wave detection \cite{revmodphysgravwave}, quantum metrology \cite{NagataSci2007,Vidrighin:2014}, quantum control \cite{GSM04} and opto-mechanical force \cite{NatNan-GavartinE-2012,PRL-harris2013} sensing has been well established.  Technological evolution in recent times has seen an increase in the range of pertinent architectures where quantum mechanical effects have become relevant \cite{PRL-harris2013,PRL-Tsang2012QPE,nphoton-Taylor2013}. There have also been experimental demonstrations of key advances in QPE. For example, in optical phase estimation we saw successive lowering of achievable mean square estimation error by the use of adaptive feedback \cite{Mabuchi} and adaptive feedback was combined with smoothing \cite{WheatleyPRL} to achieve a further reduction. With the addition of phase quadrature squeezing the limit was once more lowered \cite{science-Yonez2012}. A recent extension of these QPE techniques to a more macroscopic domain uses an optical probe beam to obtain an estimate of the position, momentum and force acting on a free-space mirror \cite{mirror}. \\
As was shown in \cite{Tsang_smooth}, an increase in estimation precision relative to filtering is expected when quantum smoothing is used. Previous work here has only considered first-order forcing noise processes with non-resonant interactions between the forcing functions and the system.  For such set-ups, results to date have yet to show a greater than two improvement of the smoothed estimate over the filtered equivalent. An interesting open question therefore remains as to whether this factor of two improvement is an upper limit for more complicated systems.  So in this work we consider a higher order forcing function that is Lorentzian in frequency. Additionally, we consider resonant interactions between the forcing function and the system (mirror) with the centre frequency of the Lorentzian aligned with the peak of a mechanical resonance. Here our theory suggests that for more realistic resonant systems driven by less benign processes, the factor of two improvement with smoothing can be improved on significantly. We present theory and simulations results showing a greater than two smoothing improvement over the equivalent optimal filtered estimate obtained. The results of the simulations both verify and extend beyond the theoretical analysis and we present experimental results to verify the simulations. 

\subsection*{Theory - optics}
To date the experimental demonstrations of smoothing have focussed on systems where the probe beam (even when quantum enhanced) interactions are in free-space. It is relatively well known that optical cavities can be used to enhance measurement precision. In the context of this work the strong intra-cavity field in an optical cavity provides more photon interaction with the parameter to be estimated. As each photon  potentially probes the parameter many times the cumulative effect gives higher sensitivity without need for extra photon resources. 
Because the experimental validation makes use of the enhancements in sensitivity achievable by the use of optical cavities, the theory and simulation assume an intra-cavity probe.\\

Using the formalisms in \cite{bachorRalph}, a single ended cavity, see figure \ref{fig:cavity}, is described in terms of optical fields as 
\bqa
\dot{a} &=&  - \kappa a + i\Delta a + \sqrt{2\kappa_{a}}A_{in} +\sqrt{2\kappa_{la}}{A_l} \label{eq:a_dot_c}
\eqa
where $A_{in}$ is the input field and $a$ is the cavity field. Here $\kappa = \kappa_a + \kappa_{la}$, $\kappa_{a}$ is the half width half maximum (HWHM) cavity decay rate and $\kappa_{la}$ is the intra-cavity loss term. We use the standard approach of separating AC and DC terms, i.e. $A = \alpha + \delta A$ and note that the loss term ($A_l$) has no DC component. After solving \eqref{eq:a_dot_c} and its conjugate for steady state ($\dot{\alpha}=0$ \& $\dot{\alpha}^* = 0$) and applying the boundary condition ($\alpha_{out} = \sqrt{2\kappa_a}\alpha - \alpha_{in}$) for the output coupler we obtain
\bqa
\alpha_{out} &=& \frac{2 \kappa_a (\kappa\alpha_{in} + i \bar{\Delta}\alpha_{in})}{\kappa^2 + \bar{\Delta}^2} - \alpha_{in}\label{eq:alphout}\\
\alpha_{out}^* &=& \frac{2 \kappa_a (\kappa\alpha_{in}^* - i \bar{\Delta}\alpha_{in}^*)}{\kappa^2 + \bar{\Delta}^2} - \alpha_{in}^*.\label{eq:alphout*}
\eqa 
\begin{figure}[!htb]
\centering{\includegraphics[height=4 cm]{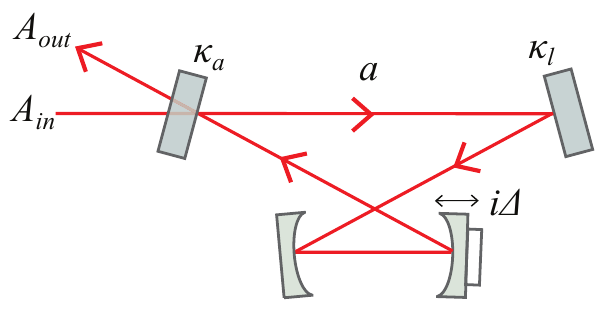}}
\caption[Cavity model]{Single ended cavity model used in derivations.} \label{fig:cavity}
\end{figure}
We use the standard quadrature definitions $X_{out}^+ = (A_{out}^\dag + A_{out})$ and $X_{out}^- = i(A_{out}^\dag - A_{out})$ and assume that $A_{in}$ is real. 
We apply an AC forcing function to a cavity mirror via a piezoelectric transducer (PZT) (bottom right figure \ref{fig:cavity}) that varies the mirror position. The goal is to estimate that forcing function with the smallest possible mean square error (MSE). The applied signal varies (or detunes) the resonant frequency of the cavity about an average value set by cavity length. This is manifested in equation \eqref{eq:a_dot_c} via the non-linear cavity detuning term $\Delta$, which is zero when the optical frequency is equal to the cavity's resonant frequency.  The result is a non zero detuning term and hence a non zero signal on the phase quadrature at the applied AC frequency. To account for this, we also separate the non-linear cavity detuning term into average and fluctuating terms giving $\Delta = \bar{\Delta} + \zeta(t)$, where $\zeta(t)$ accounts for our applied AC forcing function (see equations \eqref{eq:dA} and \eqref{eq:dAd}). The DC component of the detuning term $\bar{\Delta}$ is assumed to be zero, meaning the DC component of $X_{A_{out}}^- $ is neglected and the DC field of interest is 
\bqa 
X_{A_{out}}^+ &=& \left(\frac{2\kappa_a - \kappa}{\kappa}\right) X_{A_{in}}^+. \label{eq:para_amp}
\eqa
Now we address the fluctuating terms in \eqref{eq:a_dot_c}. We cannot assume steady state so we move to the Fourier domain to solve the differential equation. Using the relation that $\mathfrak{F}[\frac{dx}{dt}] = i\omega\tilde{x}$ and substituting the DC solutions for $\alpha$ and $\alpha^*$ as necessary we obtain
\bqa
 i\omega\tilde{\delta A}&=& - \kappa\tilde{\delta A}  + i\tilde{\zeta}\alpha  + \sqrt{2\kappa_a}  \tilde{\delta A}_{in} + \sqrt{2\kappa_l}\tilde{\delta A_l}\label{eq:dA}\\
i\omega\tilde{\delta A^{\dag}}&=& - \kappa\tilde{\delta A^{\dag}}  - i\tilde{\zeta}\alpha^*  + \sqrt{2\kappa_a} \tilde{\delta A_{in}^{\dag}} + \sqrt{2\kappa_l}\tilde{\delta A_l^{\dag}}\label{eq:dAd}.
\eqa
As we are interested in low frequencies, we assume that $\omega \ll \kappa$. After applying the boundary conditions and some relatively straight forward algebra, the output quadratures are 
\bqa
\tilde{\delta X}^-_{A_{out}} &=& \frac{(2\kappa_a-\kappa)\tilde{\delta X}^-_{A_{in}}}{\kappa} + \frac{2\sqrt{\kappa_a\kappa_l} \tilde{\delta X}^-_{A_{l}}}{\kappa} + \frac{(4\kappa_a\alpha_{in})\tilde{\zeta}}{\kappa^2} \label{eq:X-_phsq}\\
\tilde{\delta X}^+_{A_{out}} &=& \frac{(2\kappa_a-\kappa)\tilde{\delta X}^+_{A_{in}}}{\kappa} + \frac{2\sqrt{\kappa_a\kappa_l} \tilde{\delta X}^+_{A_{l}}}{\kappa}.\label{eq:X+_phsq}
\eqa
The fluctuating component of the detuning term ($\zeta$) includes the PZT response to higher frequency perturbations, i.e. the applied forcing function. It therefore makes sense to think about the output field, which is our probe of the mirror position, from a signal and noise sense. We assume that only the phase quadrature of the probe is measured. The phase quadrature (equation \eqref{eq:X-_phsq}) can been seen to consist of components that are quantum fluctuations (first two terms) and a component that is a function of the detuning parameter (last term).   The detuning ($\zeta(\omega)$) of the optical resonant frequency of the cavity is a function of the cavity length changing with mirror displacement. The mirror is coupled to the PZT and so the magnitude of the displacement depends on the PZT's frequency response. At a mechanical resonance of the PZT, a greater displacement will be imparted on the mirror for a given forcing function. Therefore $\zeta(\omega)$ is not constant with frequency but varies as a function of the PZT transfer function. By contrast $\delta X^-_{A_{in}}(\omega)$ and $\delta X^-_{A_{l}}(\omega)$ represent quantum vacuum fluctuations which are constant with frequency. \\

\subsection*{Theory - smoother}
In this subsection, we develop the theory to predict how much improvement can be expected by using smoothing to obtain our estimate compared to filtering. In the process, we will derive expressions for the optimal smoothed MSE and the transfer function of the optimal smoother. We take a block diagram approach and consider the optical part of the system as a generic \textit{plant} with input and output signals (I/O), as shown in figure \ref{fig:est_block}. At this stage it is more intuitive to consider signal voltages rather than perturbations in metres. The system definition will be recast later to derive the optimal smoother for estimating the mirror position in metres. From figure \ref{fig:est_block}, we define our system as 
\begin{figure}[!htb]
\centering{\includegraphics[height=5.4 cm]{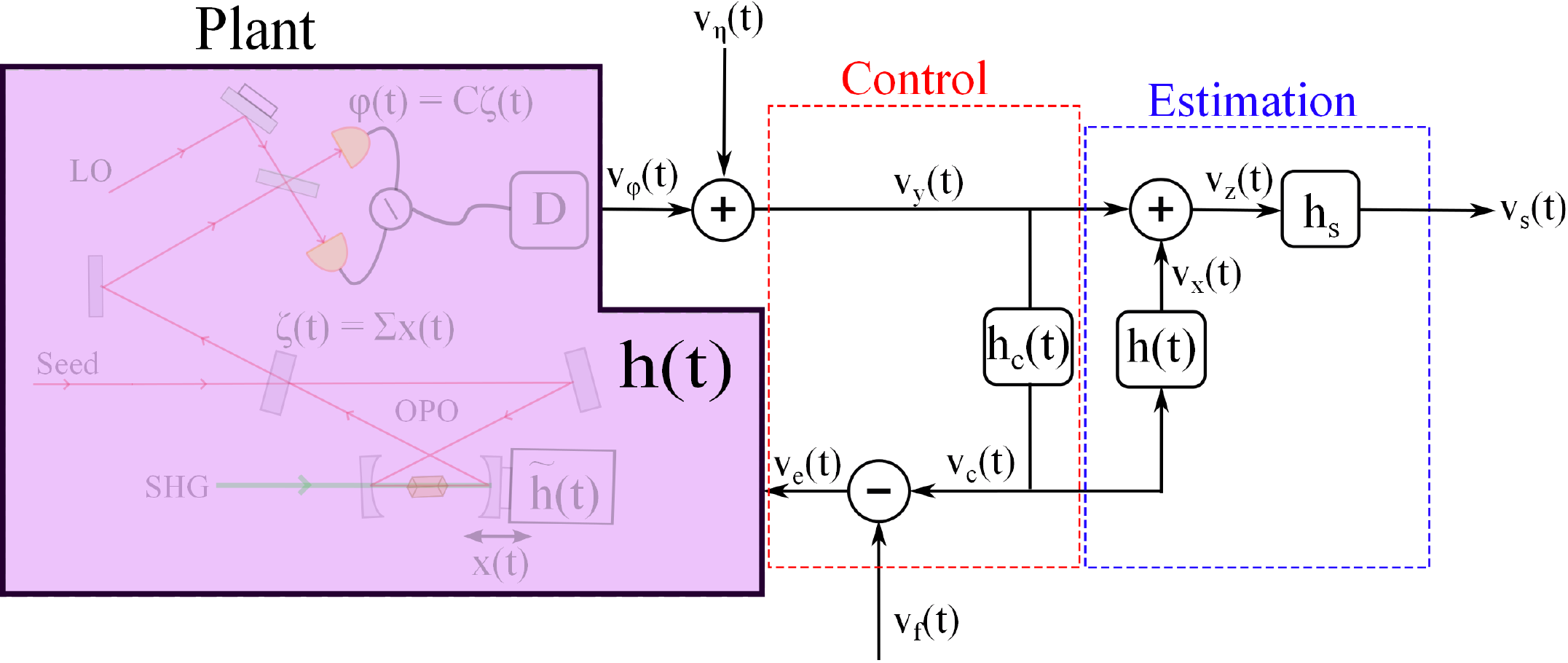}}
\caption[Voltage estimation block diagram]{The block diagram used for the derivation of the controller and the voltage estimator.} \label{fig:est_block}
\end{figure}
\bqa
v_y(t) &=& v_\varphi(t) + v_\eta(t)\label{eq:v_y}\\
\therefore v_\varphi(t) &=& h(t)*(v_f(t)-v_c(t))\label{eq:v_varp}
\eqa
where $v_\varphi(t)$ is the noiseless (unmeasurable) signal due to the plant disturbance, $v_\eta(t)$ is measurement noise (in our case dominated by quantum noise) and $v_y(t)$ is the measured output of the homodyne detector. Also $h(t)$ is the transfer function of the entire optical system, $v_f(t)$ is the applied forcing function, $v_c(t)$ is the control signal and $*$ represents a convolution operation. We now develop the optimal smoother and establish the MSE of the smoothed estimate for the system defined by \eqref{eq:v_y} and \eqref{eq:v_varp}.   Assuming that we can obtain a stable controller $h_c(t)$:
\bqa 
v_c(t) &=& h_c(t)*v_y(t)\\
\therefore v_\varphi(t) &=& h(t)*(v_f(t) - h_c(t)*[v_\varphi(t)+v_\eta(t)]).\label{eq:v_phit}
\eqa
We continue in the Fourier frequency domain where equation \eqref{eq:v_phit} becomes
\bqa
v_\varphi(\omega) &=& \frac{h(\omega)v_f(\omega)-h(\omega)h_c(\omega)v_\eta(\omega)}{1+h(\omega)h_c(\omega)}. \label{eq:v_phiw}
\eqa 
As there may be uncertainty in the system parameters, we prefer the smoothed estimate ($v_s(t)$) to be minimally impacted by a sub-optimal control signal ($v_c(t)$). This \textit{controller independence} is achieved by the addition of the $h(t)$ block in the estimator box of figure \ref{fig:est_block} so that 
\bqa 
v_z(t)
&=& h(t)*v_f(t) - {h(t)*v_c(t)} +v_\eta(t) + h(t)*v_c(t).\label{eq:z_int}
\eqa
The $v_c(t)$ terms in \eqref{eq:z_int} cancel giving
\bqa
v_z(t) &=& h(t)*v_f(t) + v_\eta(t).
\eqa
Here we note that the optimal estimation theory assumes a known transfer function $h(\omega)$, which we measured experimentally. Although this assumption suffices for our purpose, any mismatch between our assumed model and reality will lead to an increase in the actual error. Many techniques are available to address this potential problem \cite{opt_state_est}. We now derive the transfer function for the optimal smoother $h_s(\omega)$ to estimate $v_f(\omega)$,
\bqa
v_s(\omega) &=& h_s(\omega)v_z(\omega)
\eqa
and define the estimation error as
\bqa 
\Delta v(\omega) = v_s(\omega) - v_f(\omega)
= v_f(\omega)[h_s(\omega)h(\omega) - 1] + h_s(\omega)v_\eta(\omega).\label{eq:delta_v_int}
\eqa
The power spectral density of the error signal is
\bqa
S_{\Delta v}(\omega) &=&  | h_s(\omega) h(\omega) - 1 |^2 S_{v_f}(\omega) + h_s(\omega)|^2 S_{v_\eta}(\omega).\label{eq:PSD_err}
\eqa
The mean square error (MSE) is defined in the normal way and can be expressed in the frequency domain using Parseval's theorem \cite{LQG_book}
\bqa
\epsilon_v &\equiv & \mathbb{E}[\Delta v^2(t)] = \int_{-\infty}^{\infty} \frac{d\omega}{2\pi}S_{\Delta v}(\omega).\label{eq:MSE_th}
\eqa 
From equation \eqref{eq:MSE_th}, the mean square error is
\bqa
\epsilon_v &=& \int_{-\infty}^{\infty} \frac{d\omega}{2\pi}[| h_s(\omega) h(\omega) - 1 |^2 S_{v_f}(\omega) + |h_s(\omega)|^2 S_{v_\eta}(\omega)].\label{eq:mse_v}
\eqa
We minimise the MSE by finding the functional derivative of equation \eqref{eq:mse_v} with respect to $h_s(\omega)$ to obtain the optimal smoother
\bqa
h_s(\omega) &=& \frac{h^*(\omega)S_{v_\eta}(\omega)}{| h(\omega)|^2 S_{v_f}(\omega) + S_{v_\eta}(\omega)}\label{eq:hs_vf}.
\eqa
By substituting \eqref{eq:hs_vf} into \eqref{eq:mse_v}, the mean square error for the optimal estimate of $v_f(t)$ is found to be
\bqa
\epsilon_v &=& \int_{-\infty}^{\infty} \frac{d\omega}{2\pi}\frac{S_{v_f}(\omega)S_{v_\eta}(\omega)}{| h(\omega)|^2 S_{v_f}(\omega) + S_{v_\eta}(\omega)}.\label{eq:opt_mse_v}
\eqa
To recast the smoother to estimate mirror position we insert a conversion block and redefine the transfer function, see figure \ref{fig:xest_block}. It is then just a matter of reworking the above derivation with the new system. 
After reworking the algebra, we find that the optimal smoother for position estimation is
\bqa
h'_s(\omega) &=& \frac{h'^*(\omega)S_{x_f}(\omega)}{|h'(\omega)|^2 S_{x_f} + S_{v_\eta}}.\label{eq:opt_sm_x}
\eqa
where $S_{x_f} = S_{v_f}\cdot A_{_{PZT}}^2$ and $h'(\omega) = h(\omega)/A_{_{PZT}}$. From which the optimal mean square position error is found to be 
\bqa
\epsilon_x &=& \int_{-\infty}^{\infty} \frac{d\omega}{2\pi}\frac{S_{x_f}S_{v_\eta}}{|h'(\omega)|^2 S_{x_f} + S_{v_\eta}}.\label{eq:opt_err_x}
\eqa
This can be shown simply and conveniently to be  $\epsilon_x = A_{_{PZT}}^2 \epsilon_v$. Here $A_{_{PZT}}$ is a constant that relates the voltage applied to the PZT to the physical mirror displacement and was measured to be $A_{_{PZT}}\approx 6.3\times 10^{-7}$ [m/V].
\begin{figure}
\centering{\includegraphics[height=6.2 cm]{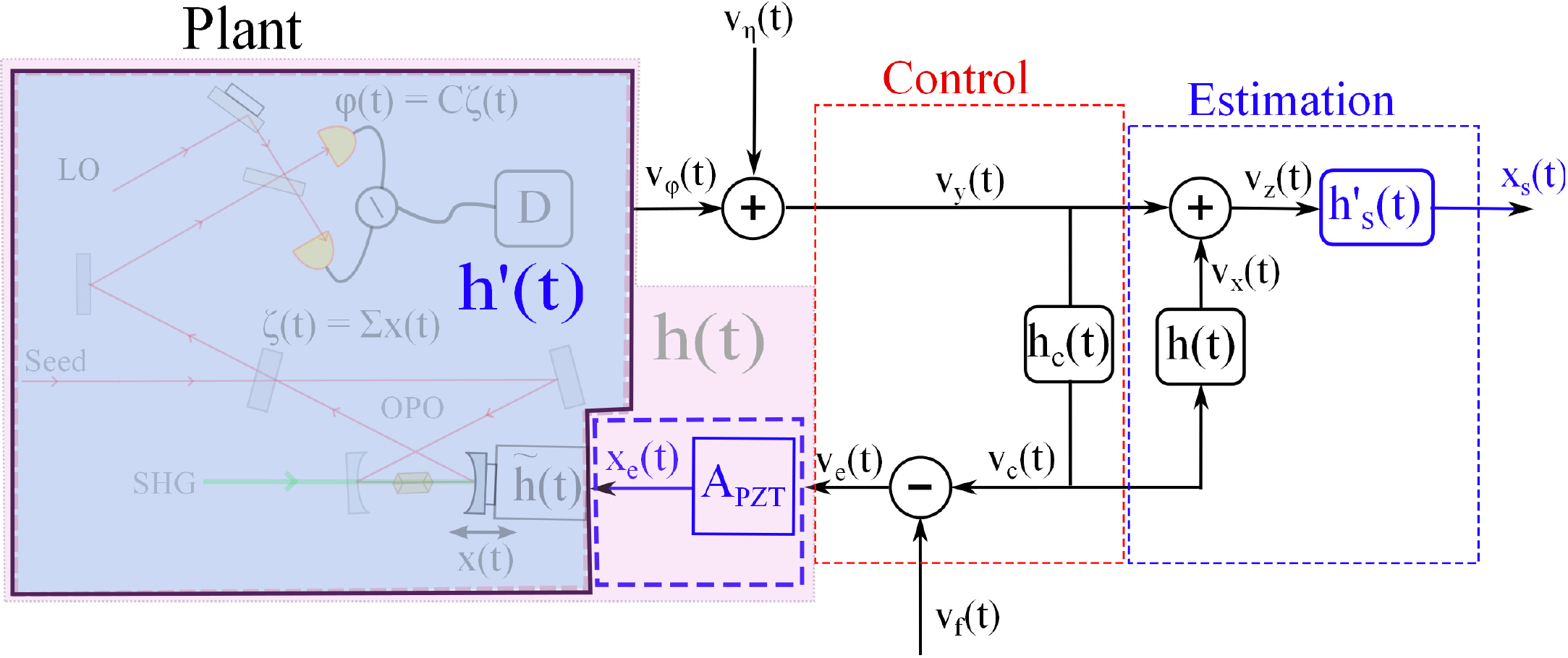}}
\caption[Mirror position estimation block diagram]{The block diagram with additional gain block used for the derivation of the position estimator. Here the transfer function $h(t)$ is modified to $h^\prime(t)$ that assumes position, $x_e(t)$ (units of metres), rather than voltage, $v_e(t)$, as an input. As such the smoothed output, $x_s(t)$, is similarly a position estimate.} \label{fig:xest_block}
\end{figure}

\subsection*{Plant and forcing function}
So far, no assumptions have been made about the dynamics of the forcing function or the plant. Of interest in this work is the MSE when those dynamics are non-trivial. In this subsection we describe both the plant and the forcing function used in the theory, simulation and later the experiment. Starting with the plant, the model used in the theory and the simulation is based on modelling a true experimental system. The transfer function of a true physical system was measured using a dynamic signal analyser. We then modelled the dominant resonance, see $h(\omega)$ in table \ref{tab:param}. The magnitude and phase plots of the measured system and the model are shown in figure \ref{fig:plant}. The plot shows that there is a time delay (identified by the constant phase lag super-imposed on the other features in the lower plot of figure \ref{fig:plant}). This delay is included in the model but not compensated for in either the controller or the smoother. It can be seen that only the dominant mechanical resonance at $\omega\approx 2\pi\times 7640 $ rad/s is accounted for in the model. The other resonances are approximately 20 dB down and are taken to be not excited by the forcing function.\\
\begin{table}[!htb]
\caption[Parameter table]{System parameters for simulation and experimental validation of the simulator.}\label{tab:param}
\centering{\begin{tabular}{|m{1.7cm}|c|c|m{3.3cm}|} 
 \hline 
 \center{\textbf{Parameter}}\vspace{4.4mm} & \textbf{Simulation} & \textbf{Experiment} & \textbf{Description} \\ 
\hline 
\vspace{0mm}  \center{$h(\omega)$} \vspace{2mm} &  \Large{$\frac{c_1s + c_2\omega_m}{s^2 + \beta s + \omega_m^2}$}\normalsize{$e^{-s\tau}$} & \Large{$\frac{c_1s + c_2\omega_m}{s^2 + \beta s + \omega_m^2}$}\normalsize{$e^{-s\tau}$} & Plant transfer function \\ 
 \hline 
\vspace{0mm}   \center{$S_f(\omega)$}\vspace{3mm} & \large{ $\frac{Q}{2}\left[\frac{1}{(\omega-\omega_i)^2 + \gamma^2} + \frac{1}{(\omega+\omega_i)^2 + \gamma^2}\right]$} &  \large{$\frac{Q}{2}\left[\frac{1}{(\omega-\omega_i)^2 + \gamma^2} + \frac{1}{(\omega+\omega_i)^2 + \gamma^2}\right]$} & Forcing function PSD\\  
 \hline 
\vspace{0mm} \center{$R$} \vspace{3mm} & $\mathrm{7.7\times 10^{-11}}$ & $\mathrm{7.7\times 10^{-11}}$ & Measurement noise magnitude term where $R\delta(t-t') = \sigma(\eta(t),\eta(t))$, $\eta(t)$ is white Gaussian noise  \\  
 \hline 
 \vspace{0mm} \center{$Q$} \vspace{3mm}&  $\mathrm{7.4\times 10^{-2}}$ & $\mathrm{7.4\times 10^{-2}}$ & Forcing function magnitude term where $Q\delta(t-t') = \sigma(\xi(t),\xi(t))$, $\xi(t)$ is white Gaussian noise \\ 
 \hline 
 \vspace{0mm} \center{$\gamma$} \vspace{3mm}& 1333 & 1333 & Forcing function damping factor\\ 
 \hline 
 \vspace{0mm} \center{$\omega_m$} \vspace{3mm} & $2\pi\cdot7930$ & $2\pi\cdot7930$ & Frequency of PZT resonance\\ 
 \hline 
 \vspace{0mm} \center{$\omega_i$}  \vspace{3mm}& $2\pi\cdot7930$ & $2\pi\cdot7930$ & Frequency of forcing function resonance\\ 
 \hline 
 \vspace{0mm} \center{$c_1$} \vspace{3mm} & 131 & 131 & Constant\\ 
 \hline 
 \vspace{0mm}  \center{$c_2$} \vspace{3mm} & 196 & 196 & Constant\\ 
 \hline 
 \vspace{0mm}  \center{$\beta$}  \vspace{3mm}& 2494 & 2494 & PZT resonance damping factor \\ 
  \hline 
\vspace{0mm} \center{$\tau$}  \vspace{3mm}& 0 and 18.5 $\times 10^{-6}$  &  system & Time delay \\
 \hline
\vspace{0mm}  \center{F} \vspace{3mm}& 250 kS/s & 250 kS/s & Sample rate\\ 
  \hline  
\vspace{0mm} \center{N} \vspace{3mm} & $2^{15}$& $2^{16}$ & Number of samples\\ 
  \hline       
\vspace{0mm} \center{Averages} \vspace{3mm}& $21$& $5$ & number of data sets averaged  \\ 
  \hline       
 \end{tabular} 
 }
 \end{table} 

\begin{figure}[!htb]
\centering{\includegraphics[width=\textwidth]{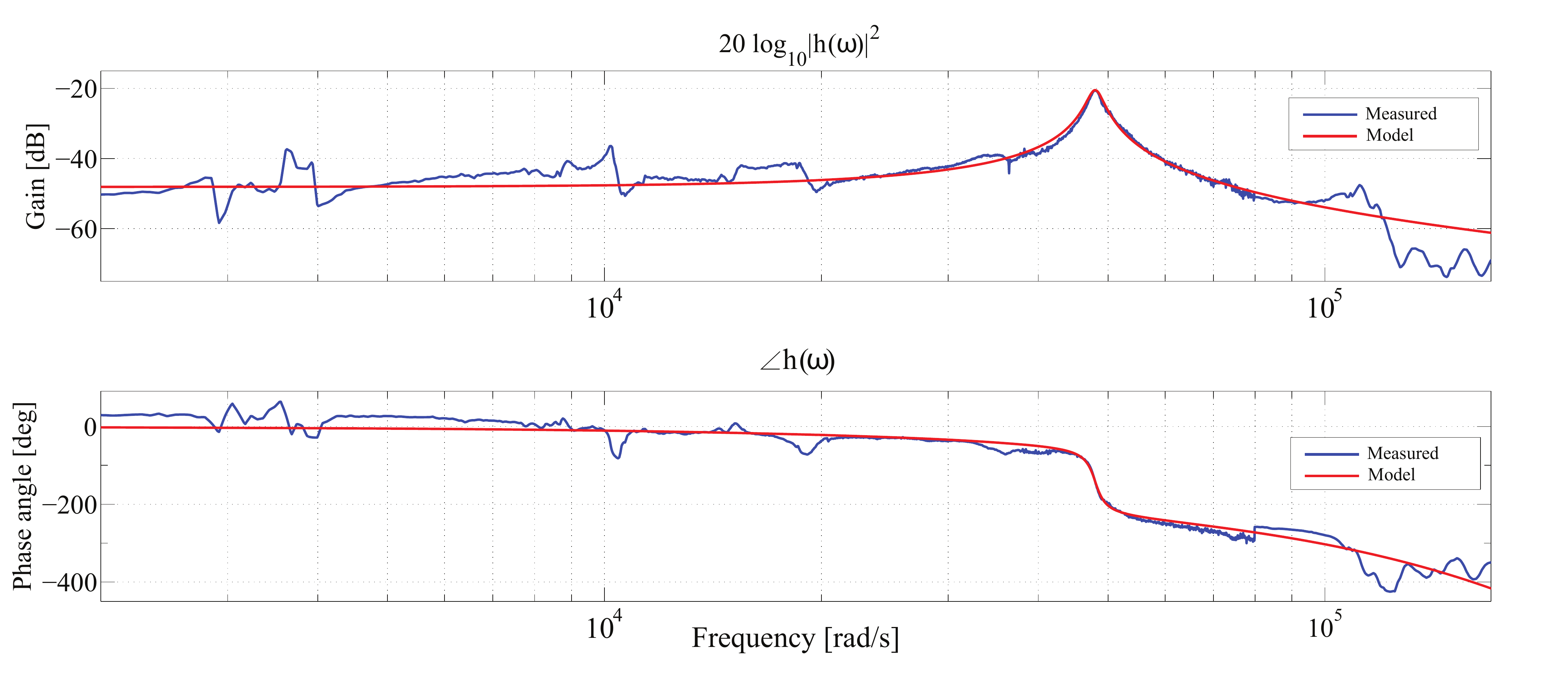}}
\caption[Bode plots of system transfer function ($h(\omega)$)]{Magnitude and phase plots of the measured system transfer function ($h(\omega)$) and the function fit used for the simulation.} \label{fig:plant}
\end{figure}
The system is driven via a cavity mirror with a Lorentzian forcing function ($v_f(t)$) as shown at figure \ref{fig:v_f}. Again, measurements of a physical system are used as the basis of the model. Figure \ref{fig:v_f} shows the PSDs of both the simulated (blue) and experimental (red) forcing functions. The model used in the controller and system design is shown in table \ref{tab:param} ($S_f(\omega)$) and is accurately represented by the blue plot in figure \ref{fig:v_f}. The experimental forcing function is also a good match to the theory for the frequency range of interest ($< 15 kHz$). It is apparent that the process used in the experiment has higher order (odd) harmonics that are not accounted for in the models. These are an artefact of the experimental generation of the forcing function. The higher order harmonics do not excite the system because both the plant and the controller are heavily attenuated at these frequencies.

\section*{Results and Discussion}

\subsection*{Simulation}
We developed a numerical simulator to test the theory and provide a baseline against which an experimental testbed can be compared. The numerical results can also be used to inform subsequent experiments. The simulation was done using \textit{Simulink} and the parameter values are shown in Table \ref{tab:param}. The input and measurement noise processes were entered as floating point arrays from the workspace.  The plant ($h(t)$) and controller ($h_c(t)$) transfer functions were implemented using transfer function blocks with the numerator and denominator coefficients extracted from the workspace. The \textit{Simulink} model provides the parameter $v_z(t)$ (see figure \ref{fig:est_block}) for the smoother. In the experimental validation discussed later a cut-down version of the \textit{Simulink} model was used to process the experimental data to obtain $v_z(t)$ from the recorded experimental values of $v_y(t)$ and $v_c(t)$.  The smoothing (both simulation and experimental) was implemented in the frequency domain using the Fourier transforms of the relevant parameters from the workspace and the \textit{Simulink} model.  The goal of the simulation was to find whether there exists a range of parameters (preferably experimentally feasible) that allow for a greater than two improvement over the optimal filtered estimate.\\
\begin{figure}[!htb]
\centering{\includegraphics[width = \textwidth]{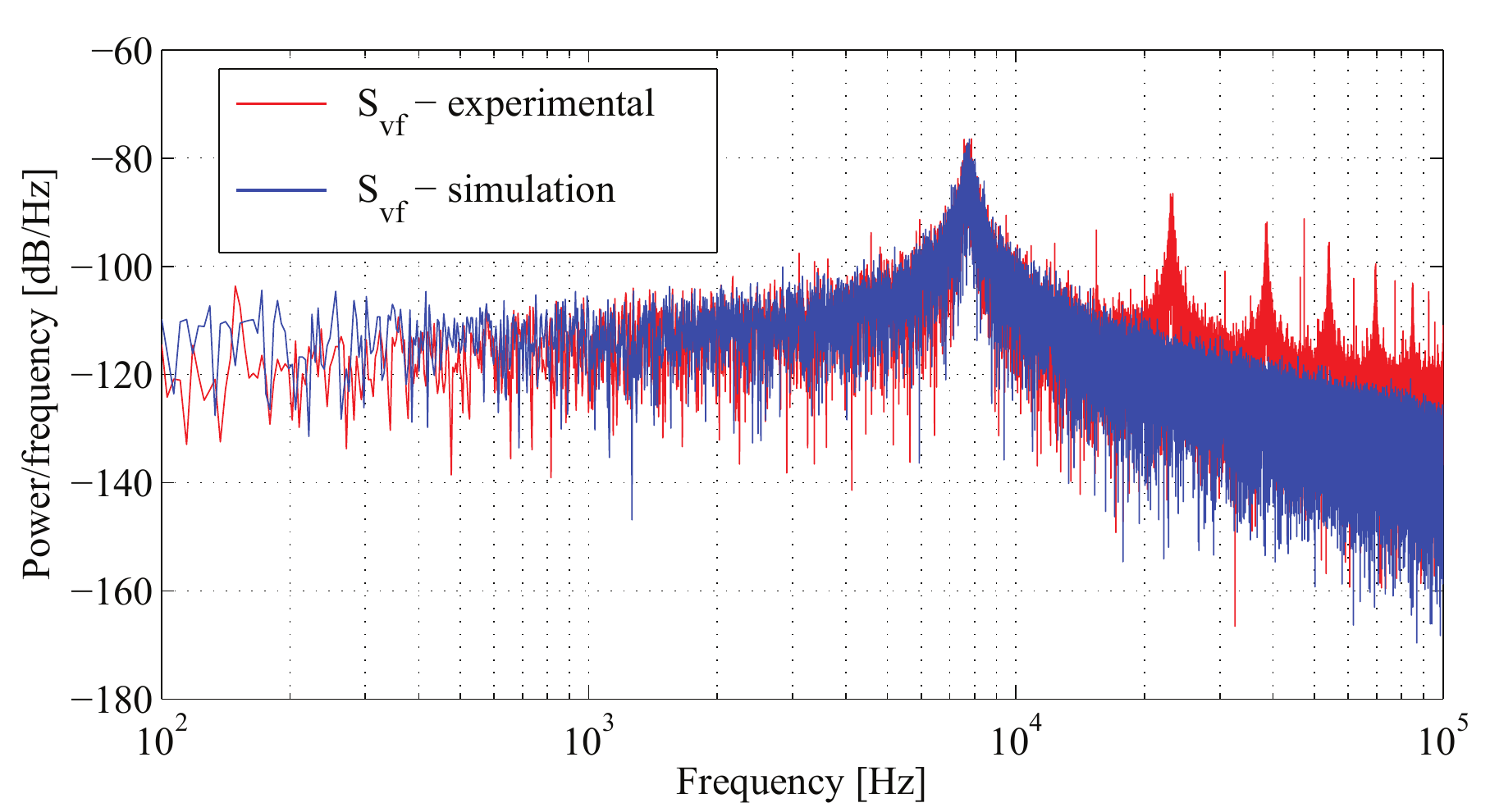}}
\caption[Experimental and simulated forcing function PSDs]{Power spectral densities of the experimental and simulated forcing functions plotted against frequency.} \label{fig:v_f}
\end{figure}
The simulation results are shown in figure \ref{fig:SIF}. This figure shows the smoothing improvement factor ($\Sigma$) for a coherent state as a function of the parameters $Q$ and $\gamma$ of the input forcing function ($S_f(\omega)$ see Table \ref{tab:param}) . The smoothing improvement factor is defined as
\bqa
\Sigma = \frac{\epsilon_{filt}}{\epsilon_{x}}, 
\eqa 
where $\epsilon_x$ is the smoothed MSE (see equation \eqref{eq:opt_err_x}) and $\epsilon_{filt}$ is the optimal filtered MSE error found numerically using the optimal Kalman-Bucy filter covariance matrix \cite{LQG_book}. The input forcing function parameter varied in the upper plot is $\gamma$, which is varied from 500 to 3082 for a fixed $Q$ of 2.35. In the lower plot we vary $Q$ from $\mathrm{9.3\times10^{-3}}$ to 2.35 for a $\gamma$ of 500. The lines are included to guide the eye between the data points. The error bars are the standard deviation of 21 separate simulations for each data point. The blue dashed line on each plot shows the theory using equation \eqref{eq:opt_err_x} and the equivalent filtered MSE for the respective parameters at the data points. The red dash-dot line shows the result for the simulation. The simulated data utilised a controller that was designed using the linear quadratic Gaussian (LQG) \cite{LQG_book} methodology with fixed central values for the $Q$ (0.57) and $\gamma$ (1193). The other parameters used in the LQG controller design are shown in Table \ref{tab:param}, with LQG design parameters  $\mu_0$ = 1 and $x_0$ = 2.4. \\ 
\begin{figure}[!htb]
\centering{\includegraphics[width = \textwidth]{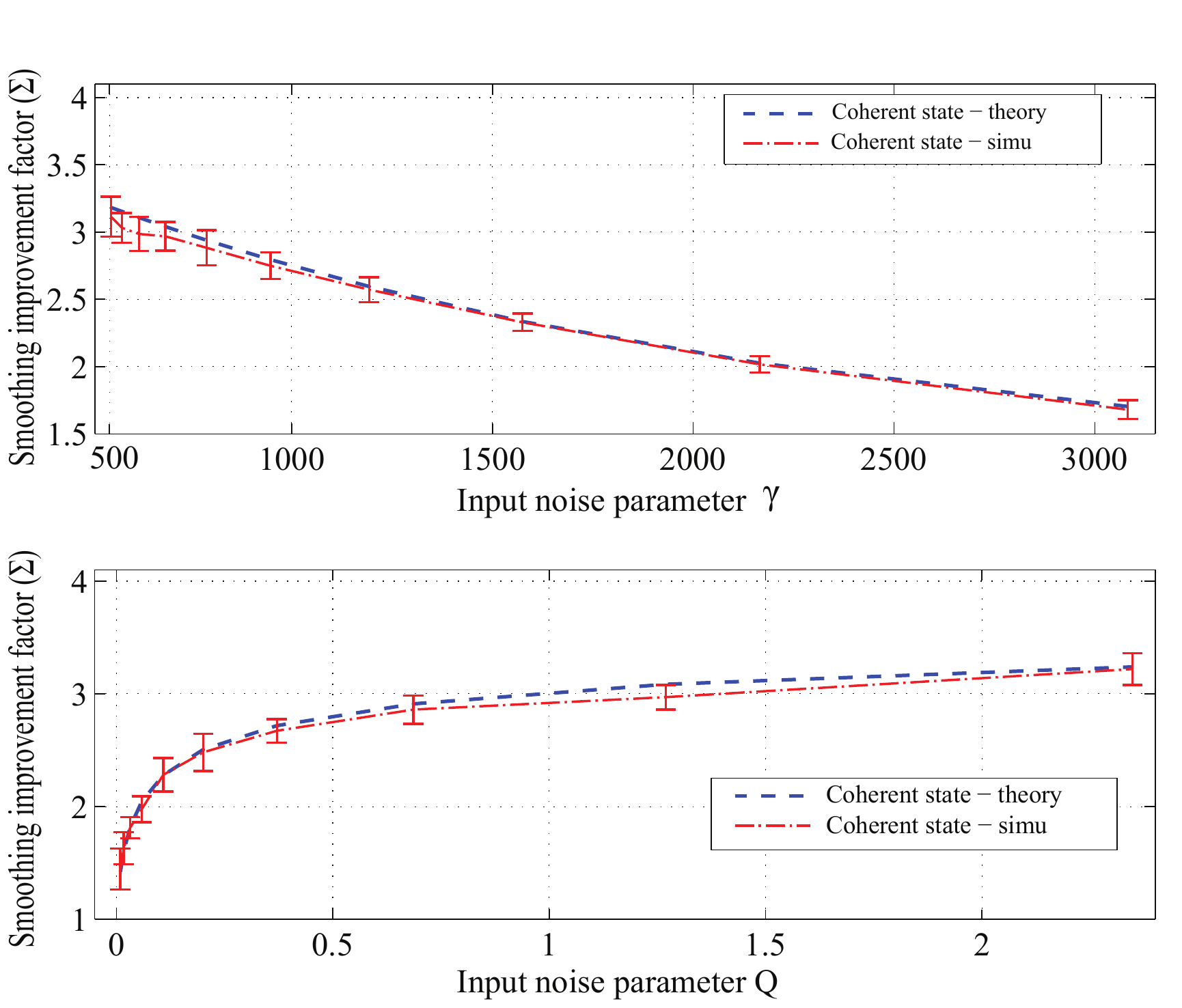}}
\caption[Parameter space vary showing Smoothing improvement factor as a factor of $Q$ and $\gamma$]{Smoothing improvement factor ($\Sigma$) for a coherent state for varying input forcing function ($S_f(\omega)$  parameters $Q$ and $\gamma$.} \label{fig:SIF}
\end{figure}
It is quite clear in figure \ref{fig:SIF} that there is good agreement between simulation and theory, with most data points agreeing within error bars. It is noted that the model assumes that the optical cavity remains linear and as such does not account for large values of the non-linear detuning term ($\Delta$ in equation \eqref{eq:a_dot_c}).  The theory and simulation show that a greater than two smoothing enhancement for a resonant process acting on a mechanically resonant structure is achievable for a wide range of parameters.
 \subsection*{Experiment}
 The goal of the experimental results presented in this paper are to validate that the theory and simulation results appropriately reflect a physically reasonable experimental system.  The experimental set up is shown in figure \ref{fig:exp}. We use the 1064 nm output of an Innolight ``Diabolo'' doubled NdYAG laser as the primary optical frequency for the cavity. The 1064 nm beam is spatially filtered using an MCC. After the MCC, 35.6 mW is split off at a 100:1 ratio with a polarising beam splitter (PBS) for use as the local oscillator (LO) for balanced homodyne measurement. The remaining light of approximately 400 $\mathrm{\mu W}$ is phase modulated at 199 MHz (RF1 on figure \ref{fig:exp}) to create a weak coherent state ($ n \approx 10^3$ photons per second). This modulated coherent state is used as the input to a single ended bow-tie cavity, with an FSR of 199 MHz \cite{Siegman}. The cavity is locked using dither locking \cite{PDH,EB01} with a frequency of 1.322047 MHz (RF2 on figure \ref{fig:exp}) applied through the EOM and detected at PD3 on figure \ref{fig:exp}. This dither signal is demodulated and a low frequency PI controller is used to maintain the DC frequency locking of the cavity ($\bar{\Delta}$ in equation \eqref{eq:a_dot_c}).  The cavity output (Sig) is sent to a spatial balanced homodyne detector with a fringe visibility of 95.9\% and 97.9\% for each detector respectively (averaged in the quantum efficiency calculation). The low frequency (LF) output of the homodyne detector via a PI controller is used to lock the detection to the phase quadrature of the signal. The high frequency (HF) homodyne output is demodulated using a 199 MHz (RF1) radio frequency (RF) LO, an RF mixer (Mix1) and a low pass filter (LPF). This demodulated signal is then feed into the feedback filter (FBF). The input signal to the FBF ($v_y(t)$) is stored using an Acqiris data acquisition system with a sampling rate (F) of 250 kS/s for post processing.  The output of the FBF ($v^\prime _c(t)$) is captured before an attenuator (Attn) for better signal to noise ratio and is also stored for post processing. The output of the attenuator ($v_c(t)$) is added to the DC lock signal and the applied forcing function (Noise). It is then amplified by a high voltage amplifier (HV amp) and applied to the PZT attached to the cavity mirror to be estimated. The forcing function signal ($S_{f}(\omega)$ in Table \ref{tab:param} and also figure \ref{fig:v_f}) is generated by amplitude modulating an Ornstein-Uhlenbeck (OU) process generated with an operational amplifier circuit and a white noise generator with a carrier of frequency $\omega_c$.  This signal ($v_f(t)$) is stored for comparison with the estimation and applied to the cavity mirror PZT. The signal $v_f(t)$ imposes the variation of mirror position that is to be estimated and is the reference for calculation of the MSE. As the data acquisition system has a 50 $\Omega$ input impedance all acquired signals are buffered with unity gain operational amplifier circuits so that the acquisition has minimal effect on the voltage levels. Additionally anti-aliasing filters were used on all channels of the data acquisition system and the laser's resonant relaxation oscillation was suppressed by its noise eater.\\
\begin{figure}
\centering{\includegraphics[width=\textwidth]{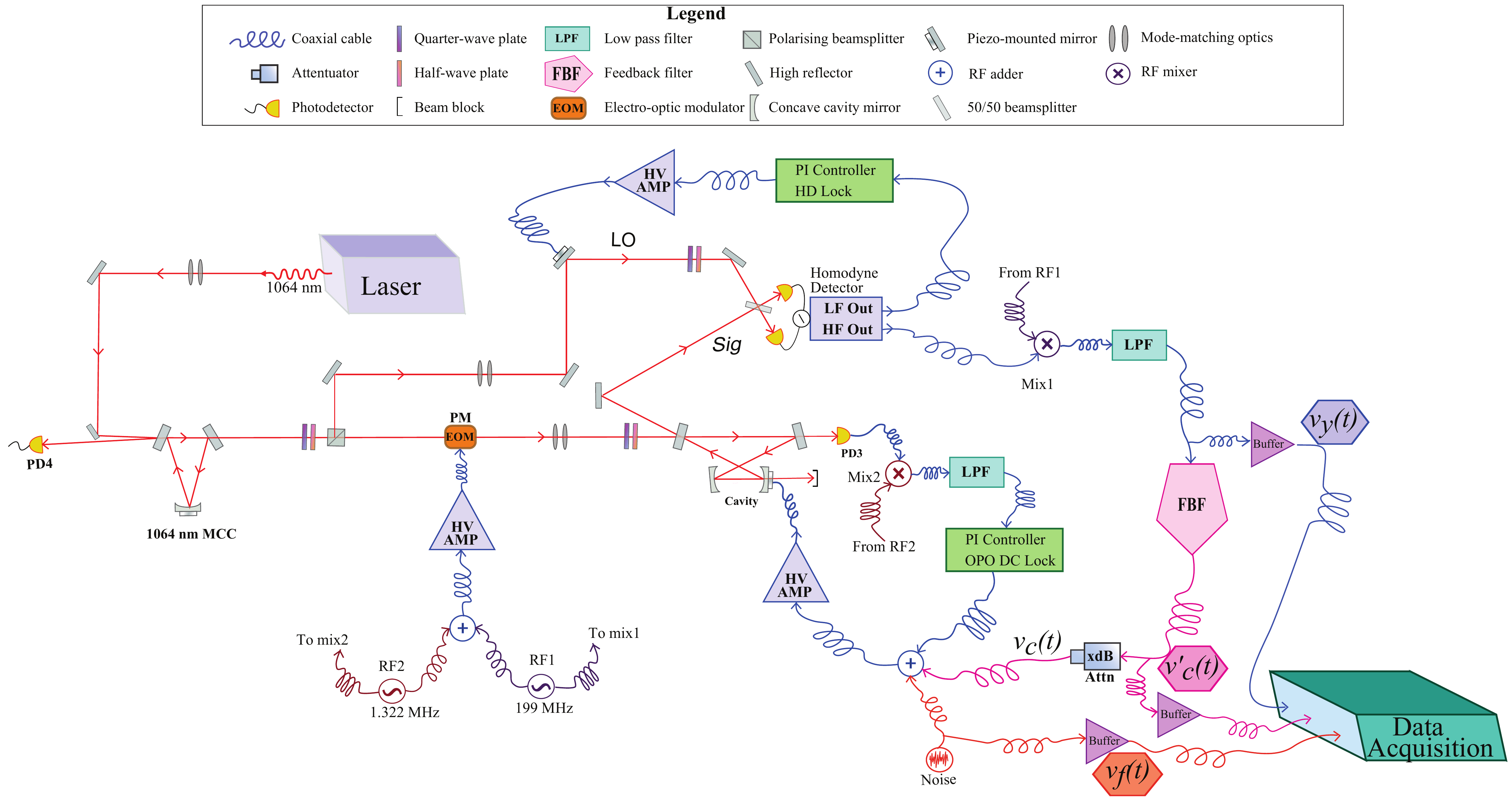}}
\caption[Full experimental setup]{The full experimental setup for the coherent state cavity mirror position estimation. Note mode cleaning cavity (MCC) locking circuitry is omitted.} \label{fig:exp}
\end{figure} 

\begin{figure}[!htb]
\centering{\includegraphics[width = \textwidth]{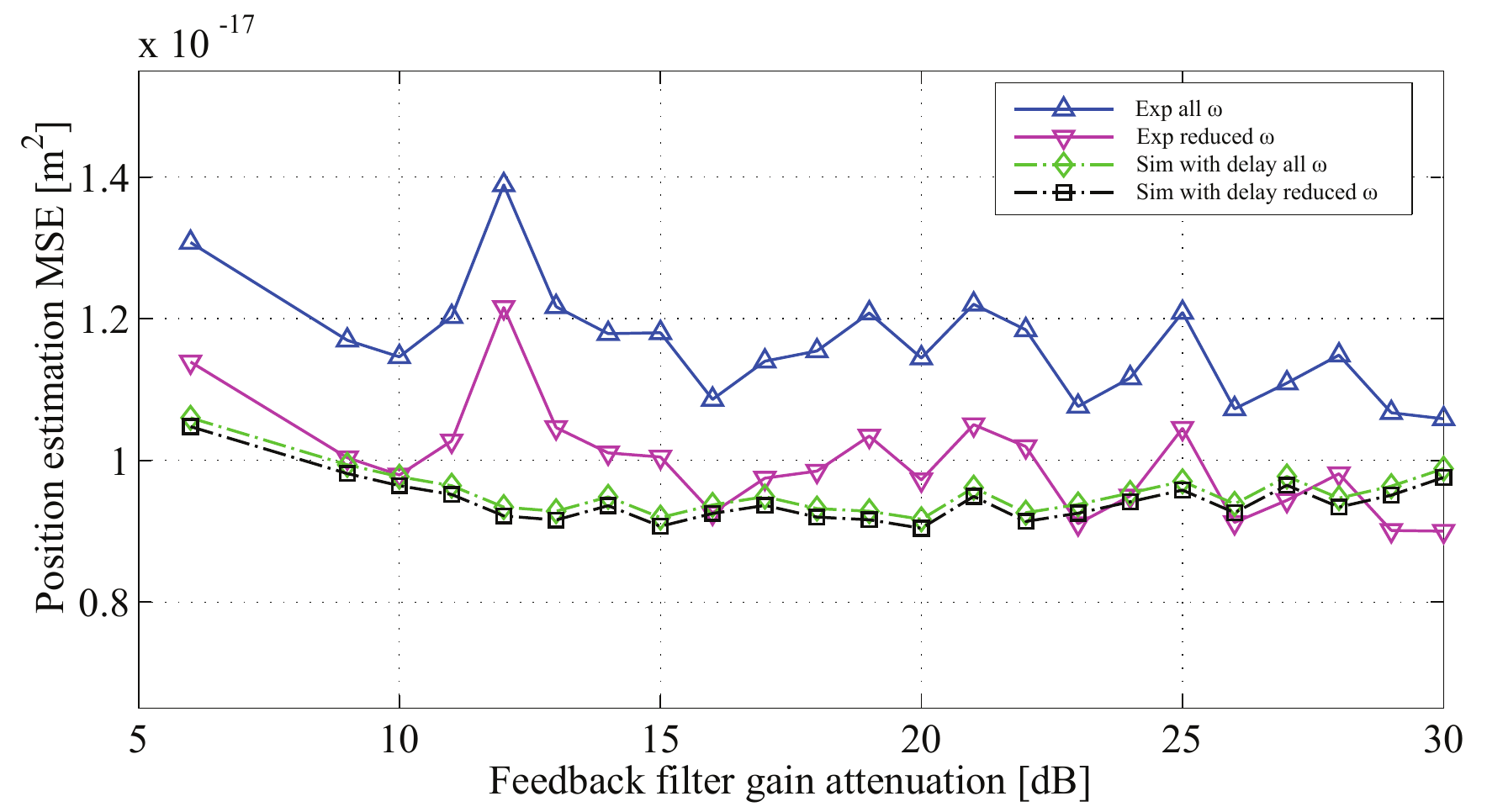}}
\caption[Experiment and simulated MSEs]{The actual MSEs (blue triangles and green diamonds) and the frequency truncated MSEs (pink inverted triangles and black squares) in position for both experiment and simulator as a function of feedback gain.} \label{fig:expMSE_ho}
\end{figure}
In the previous sub-section, we have presented the results of the theory and simulation. These results show the predicted smoothing enhancement consistent with theory and now we use the experimental data as validation of the theory.  The measured experimental parameters are summarised and compared to those of the simulation in Table \ref{tab:param}. A controller was constructed using analogue electronics and standard controller design techniques \cite{lin_cont_sys}. Whilst the design process suggested the controller was stable, the control input was also variably attenuated for additional safety. The MSE results presented here (see figure \ref{fig:expMSE_ho}, lines included to guide the eye) are plotted as a function of this variable attenuation.   There are two significant technical differences between the experiment and the simulation that need to be considered. The first  mentioned earlier is the time delay, this is simply accounted for in the \textit{Simulink} model with a delay block. The second, also mentioned earlier (see figure \ref{fig:v_f}), is the higher order harmonics on the captured forcing function. These harmonics are below the system noise floor so their effects are negligible in terms of the estimation. As they are not modelled in either the theory or the simulation the experimental data was corrected as follows and in figure \ref{fig:expMSE_ho} for the validation. To obtain the actual MSE from equation \eqref{eq:opt_err_x} the integration limits are infinite. In practice this is impossible due to finite sampling rates. So the integration limits were set at $\pm125$ kHz limited by our 250 kS/s sampling rate. The contribution of frequencies greater than 125 kHz was found to be of the order of 1\% for this system. The 125 kHz limit is thus considered reasonable for the theory and simulation comparisons, but it includes the harmonics in the experimental case. The experimental MSE is artificially inflated relative to the simulation where the forcing function has no harmonics, shown by the separation of the experimental (blue triangles) and the simulated (green diamonds) MSEs in figure \ref{fig:expMSE_ho}. To correct and allow for a fair comparison, we truncate the integration range to $\pm15$ kHz for both. The small separation of the MSE (green diamonds) and the reduced frequency MSE (black squares) in figure \ref{fig:expMSE_ho} shows that for the simulation, this truncation has only a minor impact. However, due to the removal of the unmodelled harmonics, the impact of the truncation is much greater in the experimental MSE (pink inverted triangles). The reduced frequency MSEs include only modelled data and so are suitable for the experimental validation of the simulation. Figure \ref{fig:expMSE_ho} shows that the corrected MSEs for the simulation (black squares) and the experiment (pink inverted triangles) are consistent, thus validating the simulation. The MSEs also show a degree of independence from the control input (horizontal with varied control gain) as a result of the deliberate cancellation of the control signal dependence in the smoother design. Controller independence is likely to be a useful feature in situations where uncertainty in the system model exists and will be further investigated in future work.\\


\begin{figure}[!htb]
\centering{\includegraphics[width = \textwidth]{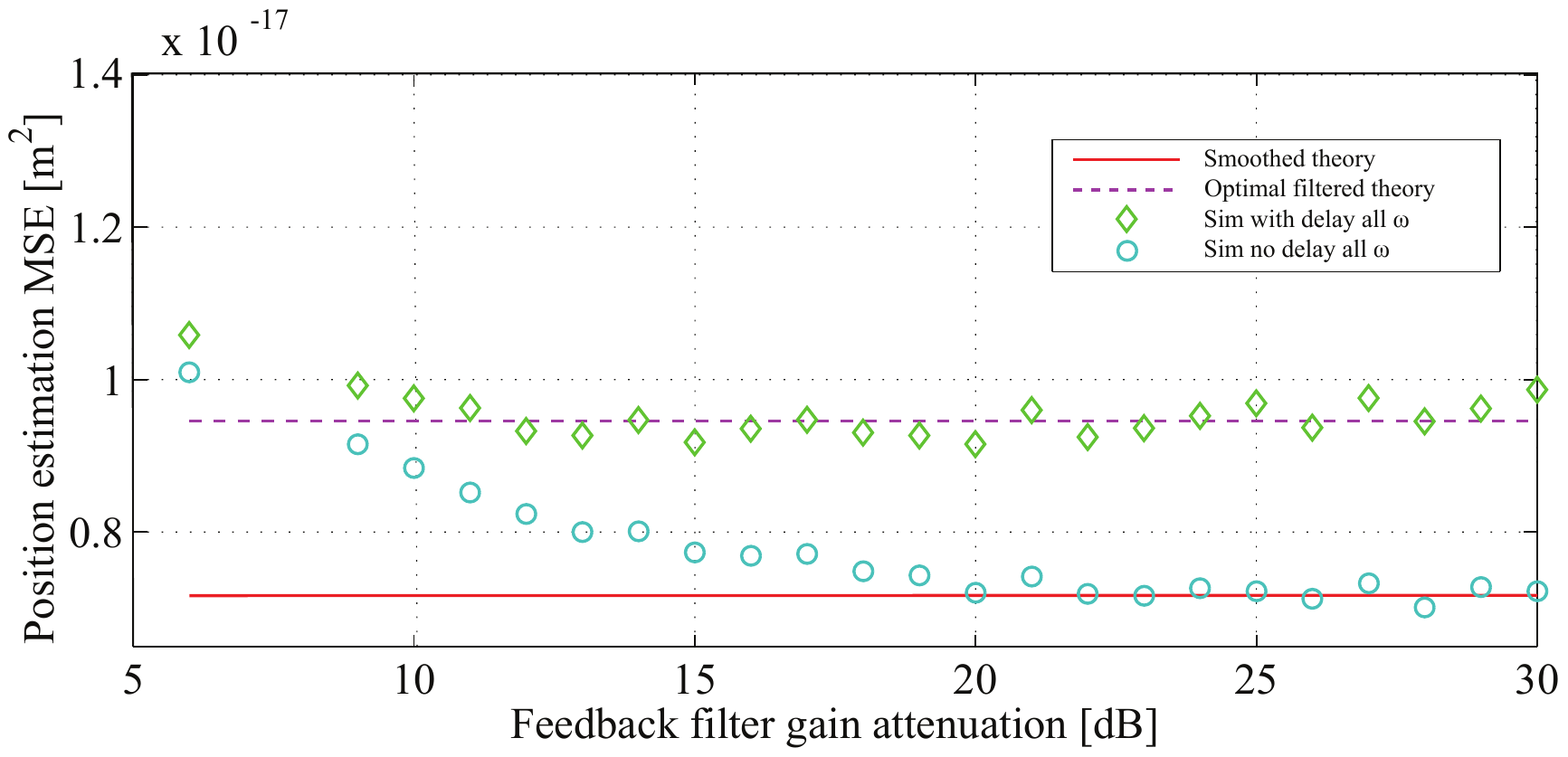}}
\caption[Experiment and simulated MSEs]{Mean square position error for both experiment and simulator as a function of feedback gain.} \label{fig:expMSE_val}
\end{figure}
 
Finally we use the experimentally validated simulator to confirm the theory in the current experimental parameter regime. Figure \ref{fig:expMSE_val} (error bars omitted for ease of viewing) shows the MSE for the simulation (green diamonds) as compared to the theoretical smoothed MSE (red solid). The offset can be explained by the fact that the smoother does not compensate for the time delay. By removing the time delay block from the simulator (cyan circles) this offset is removed and the simulated MSEs become consistent with the smoothed theory.  There is a gradual decline in precision for smaller attenuation values for both the experimental and the simulation results (see figure \ref{fig:expMSE_ho}), this may be due to controller sub-optimally starting to make the system go unstable. The final data in this plot is the MSE optimal filtered estimate (dashed purple) which is included for comparison. It is evident that the smoothed estimate is comparable with but not better than the system with time delay. However, the no time delay estimate is clearly better than the filtered equivalent but has a smoothing improvement factor of less than two but this is expected in this parameter space.  One final point of interest is in the actual value of the MSE achieved which is approximately $\mathrm{1\times 10^{-17}}$ $\mathrm{m^2}$ with $|\alpha|^2 \approx 10^3$ $s^{-1}$. This is comparable with the position MSE achieved in \cite{mirror} with squeezing enhancement and $|\alpha|^2 \approx 10^7$ $s^{-1}$. Whilst this result in itself is not surprising as it is known that optical cavities provide additional sensitivity, it is still a good outcome. With future improvements to the system, we expect to see further significant lowering of this coherent state MSE.

 \section*{Conclusions}
We have developed theory describing resonance enhanced mirror position estimation of a cavity mirror using quantum smoothing. This theory has been used to design a numerical simulation model, which we have experimentally validated. We have demonstrated that performing quantum smoothing on a mechanically resonant structure when driven by a resonant forcing function gives greater enhancement in precision when compared to non-resonant systems. When driven by a Lorentzian process we achieved a simulated improvement in precision of greater than two times better than the equivalent optimal filter, which is consistent with theory. We have also experimentally validated the simulation using an experimental testbed. The simulations have identified a good parameter regime where greater improvement should be possible in future experiments. With future improvements in the system we expect to see further precision enhancements. In future work it should be possible to demonstrate further improvement in precision by the incorporation of quantum enhancement using a phase squeezed probe beam. These results demonstrate the advantage of resonances when performing quantum parameter estimation. This is an initial proof of concept that may have applications in areas where mechanical systems are being measured in quantum limited domains.  \\

%
%
%

\section*{Competing interests}
 The authors declare that they have no competing interests.

\section*{Authors contributions}
The idea for this work and the initial smoother design came from MT in collaboration with EHH. The experimental design and theory development was done by TAW, MT and EHH. The starting LQG controller design was MT’s and it was developed and implemented by TAW, EHH and IRP. TAW developed the simulator, built the experiment and collected the data with input from EHH. The data analysis was done by TAW in consultation with EHH, IRP and MT. The paper was drafted by TAW with significant guidance from EHH and revision by IRP.

\section*{Acknowledgements}
This work was supported financially by the Australian Research Council, Grant No. CE110001029,  DP1094650, FL110100020 and DP109465.\\
Mankei Tsang acknowledges support from the Singapore National Research Foundation under NRF Grant No. NRF-NRFF2011-07.\\

{\ifthenelse{\boolean{publ}}{\footnotesize}{\small}
\bibliographystyle{bmc_article} 
\bibliography{Wheatley_et_al_cavity_bib}      


\begin{thebibliography}{10}
\providecommand{\url}[1]{[#1]}
\providecommand{\urlprefix}{}

\bibitem{PRA_bollinger1996}
Bollinger JJ, Itano WM, Wineland DJ, Heinzen DJ: \textbf{Optimal frequency
  measurements with maximally correlated states}. \emph{Phys. Rev. A} 1996,
  \textbf{54}:R4649--R4652,
  \urlprefix\url{[http://link.aps.org/doi/10.1103/PhysRevA.54.R4649]}.

\bibitem{Pryde}
Higgins BL, Berry DW, Bartlett SD, Wiseman HM, Pryde GJ:
  \textbf{Entanglement-free Heisenberg-limited phase estimation}. \emph{Nature}
  2007, \textbf{450}:393--396.

\bibitem{Nat-Breit1997}
Breitenbach G, Schiller S, Mlynek J: \textbf{Measurement of the quantum states
  of squeezed light}. \emph{Nature} 1997, \textbf{387}(6632):471--475.

\bibitem{wiseman_PRL_75}
Wiseman HM: \textbf{Adaptive Phase Measurements of Optical Modes: Going Beyond
  the Marginal Q Distribution}. \emph{Phys. Rev. Lett.} 1995,
  \textbf{75}(25):4587.

\bibitem{Tsang_smooth}
Tsang M: \textbf{Time-symmetric quantum theory of smoothing}. \emph{Phys. Rev.
  Lett.} 2009, \textbf{102}:250403.

\bibitem{revmodphysgravwave}
Adhikari RX: \textbf{Gravitational radiation detection with laser
  interferometry}. \emph{Rev. Mod. Phys.} 2014, \textbf{86}:121--151,
  \urlprefix\url{[http://link.aps.org/doi/10.1103/RevModPhys.86.121]}.

\bibitem{NagataSci2007}
Nagata T, Okamoto R, O'Brien JL, Sasaki K, Takeuchi S: \textbf{Beating the
  Standard Quantum Limit with Four-Entangled Photons}. \emph{Science} 2007,
  \textbf{316}(5825):726--729,
  \urlprefix\url{[http://www.sciencemag.org/content/316/5825/726.abstract]}.

\bibitem{Vidrighin:2014}
Vidrighin MD, Donati G, Genoni MG, Jin XM, Kolthammer WS, Kim MS, Datta A,
  Barbieri M, Walmsley IA: \textbf{Joint estimation of phase and phase
  diffusion for quantum metrology}. \emph{Nat Commun} 2014, \textbf{5},
  \urlprefix\url{[http://dx.doi.org/10.1038/ncomms4532]}.

\bibitem{GSM04}
Geremia J, Stockton J, Mabuchi H: \textbf{Real-time quantum feedback control of
  atomic spin-squeezing}. \emph{Science} 2004, (304):270--273.

\bibitem{NatNan-GavartinE-2012}
Gavartin E, Verlot P, Kippenberg TJ: \textbf{A hybrid on-chip optomechanical
  transducer for ultrasensitive force measurements}. \emph{Nat Nano} 2012,
  \textbf{7}(8):509--514,
  \urlprefix\url{[http://dx.doi.org/10.1038/nnano.2012.97]}.

\bibitem{PRL-harris2013}
Harris GI, McAuslan DL, Stace TM, Doherty AC, Bowen WP: \textbf{Minimum
  Requirements for Feedback Enhanced Force Sensing}. \emph{Phys. Rev. Lett.}
  2013, \textbf{111}:103603,
  \urlprefix\url{[http://link.aps.org/doi/10.1103/PhysRevLett.111.103603]}.

\bibitem{PRL-Tsang2012QPE}
Tsang M: \textbf{Ziv-Zakai Error Bounds for Quantum Parameter Estimation}.
  \emph{Phys. Rev. Lett.} 2012, \textbf{108}:230401,
  \urlprefix\url{[http://link.aps.org/doi/10.1103/PhysRevLett.108.230401]}.

\bibitem{nphoton-Taylor2013}
Taylor MA, Janousek J, Daria V, Knittel J, Hage B, BachorHans-A, Bowen WP:
  \textbf{Biological measurement beyond the quantum limit}. \emph{Nat Photon}
  2013, \textbf{7}(3):229--233,
  \urlprefix\url{[http://dx.doi.org/10.1038/nphoton.2012.346]}.

\bibitem{Mabuchi}
Armen M, Au J, Stockton J, Doherty A, Mabuchi H: \textbf{Adaptive homodyne
  measurement of optical phase}. \emph{Phys. Rev. Lett.} 2002,
  \textbf{89}:133602.

\bibitem{WheatleyPRL}
Wheatley T, Berry D, Yonezawa H, Nakane D, Arao H, Pope D, Ralph T, Wiseman H,
  Furusawa A, Huntington E: \textbf{Adaptive Optical Phase Estimation Using
  Time-Symmetric Quantum Smoothing}. \emph{Phys. Rev. Lett.} 2010,
  \textbf{104}(093601).

\bibitem{science-Yonez2012}
Yonezawa H, Nakane D, Wheatley TA, Iwasawa K, Takeda S, Arao H, Ohki K, Tsumura
  K, Berry DW, Ralph TC, Wiseman HM, Huntington EH, Furusawa A:
  \textbf{Quantum-Enhanced Optical-Phase Tracking}. \emph{Science} 2012,
  \textbf{337}(6101):1514--1517,
  \urlprefix\url{[http://www.sciencemag.org/content/337/6101/1514.abstract]}.

\bibitem{mirror}
Iwasawa K, Makino K, Yonezawa H, Tsang M, Davidovic A, Huntington E, Furusawa
  A: \textbf{Quantum-Limited Mirror-Motion Estimation}. \emph{Phys. Rev. Lett.}
  2013, \textbf{111}:163602,
  \urlprefix\url{[http://link.aps.org/doi/10.1103/PhysRevLett.111.163602]}.

\bibitem{bachorRalph}
Bachor HA, Ralph TC: \emph{A Guide to Experiments in Quantum Optics}. Weinheim:
  Wiley-VCH, 2nd edition 2004.

\bibitem{opt_state_est}
Simon D: \emph{Optimal State Estimation: Kalman, H Infinity, and Nonlinear
  Approaches}. Hoboken: Wiley 2006,
  \urlprefix\url{[http://books.google.com.sg/books?id=urhgTdd8bNUC]}.

\bibitem{LQG_book}
Dorato P, Abdallah CT, Cerone V: \emph{Linear Quadratic Control : An
  Introduction}. New York: MacMillan 1995.

\bibitem{Siegman}
Siegman AE: \emph{Lasers}. University Science Books 1986.

\bibitem{PDH}
Drever RWP, Hall JL, Kowalski FV, Hough J, Ford GM, Munley AJ, Ward H:
  \textbf{Laser Phase and Frequency Stabilization Using an Optical Resonator}.
  \emph{Appl. Phys. B} 1983, \textbf{31}:97--105.

\bibitem{EB01}
Black E: \textbf{An introduction to {P}ound--{D}rever--{H}all laser frequency
  stabilization}. \emph{Am. J. Phys.} 2001, \textbf{69}:79--87.

\bibitem{lin_cont_sys}
Rohrs CE, Melsa JL, Schultz DG: \emph{Linear Control Systems}. Singapore:
  McGraw-Hill, international edition edition 1993.

\end{thebibliography}

\newcommand{\BMCxmlcomment}[1]{}

\BMCxmlcomment{

<refgrp>

<bibl id="B1">
  <title><p>Optimal frequency measurements with maximally correlated
  states</p></title>
  <aug>
    <au><snm>Bollinger</snm><fnm>J. J .</fnm></au>
    <au><snm>Itano</snm><fnm>WM</fnm></au>
    <au><snm>Wineland</snm><fnm>D. J.</fnm></au>
    <au><snm>Heinzen</snm><fnm>D. J.</fnm></au>
  </aug>
  <source>Phys. Rev. A</source>
  <publisher>American Physical Society</publisher>
  <pubdate>1996</pubdate>
  <volume>54</volume>
  <fpage>R4649</fpage>
  <lpage>-R4652</lpage>
  <url>http://link.aps.org/doi/10.1103/PhysRevA.54.R4649</url>
</bibl>

<bibl id="B2">
  <title><p>Entanglement-free Heisenberg-limited phase estimation</p></title>
  <aug>
    <au><snm>Higgins</snm><fnm>B. L.</fnm></au>
    <au><snm>Berry</snm><fnm>D. W.</fnm></au>
    <au><snm>Bartlett</snm><fnm>S. D.</fnm></au>
    <au><snm>Wiseman</snm><fnm>H. M.</fnm></au>
    <au><snm>Pryde</snm><fnm>G. J.</fnm></au>
  </aug>
  <source>Nature</source>
  <pubdate>2007</pubdate>
  <volume>450</volume>
  <fpage>393</fpage>
  <lpage>396</lpage>
</bibl>

<bibl id="B3">
  <title><p>Measurement of the quantum states of squeezed light</p></title>
  <aug>
    <au><snm>Breitenbach</snm><fnm>G</fnm></au>
    <au><snm>Schiller</snm><fnm>S</fnm></au>
    <au><snm>Mlynek</snm><fnm>J</fnm></au>
  </aug>
  <source>Nature</source>
  <pubdate>1997</pubdate>
  <volume>387</volume>
  <issue>6632</issue>
  <fpage>471</fpage>
  <lpage>-475</lpage>
</bibl>

<bibl id="B4">
  <title><p>Adaptive Phase Measurements of Optical Modes: Going Beyond the
  Marginal Q Distribution</p></title>
  <aug>
    <au><snm>Wiseman</snm><fnm>H. M.</fnm></au>
  </aug>
  <source>Phys. Rev. Lett.</source>
  <pubdate>1995</pubdate>
  <volume>75</volume>
  <issue>25</issue>
  <fpage>4587</fpage>
</bibl>

<bibl id="B5">
  <title><p>Time-symmetric quantum theory of smoothing</p></title>
  <aug>
    <au><snm>Tsang</snm><fnm>M.</fnm></au>
  </aug>
  <source>Phys. Rev. Lett.</source>
  <pubdate>2009</pubdate>
  <volume>102</volume>
  <fpage>250403</fpage>
</bibl>

<bibl id="B6">
  <title><p>Gravitational radiation detection with laser
  interferometry</p></title>
  <aug>
    <au><snm>Adhikari</snm><fnm>RX</fnm></au>
  </aug>
  <source>Rev. Mod. Phys.</source>
  <publisher>American Physical Society</publisher>
  <pubdate>2014</pubdate>
  <volume>86</volume>
  <fpage>121</fpage>
  <lpage>-151</lpage>
  <url>http://link.aps.org/doi/10.1103/RevModPhys.86.121</url>
</bibl>

<bibl id="B7">
  <title><p>Beating the Standard Quantum Limit with Four-Entangled
  Photons</p></title>
  <aug>
    <au><snm>Nagata</snm><fnm>T</fnm></au>
    <au><snm>Okamoto</snm><fnm>R</fnm></au>
    <au><snm>O'Brien</snm><fnm>JL</fnm></au>
    <au><snm>Sasaki</snm><fnm>K</fnm></au>
    <au><snm>Takeuchi</snm><fnm>S</fnm></au>
  </aug>
  <source>Science</source>
  <pubdate>2007</pubdate>
  <volume>316</volume>
  <issue>5825</issue>
  <fpage>726</fpage>
  <lpage>729</lpage>
  <url>http://www.sciencemag.org/content/316/5825/726.abstract</url>
</bibl>

<bibl id="B8">
  <title><p>Joint estimation of phase and phase diffusion for quantum
  metrology</p></title>
  <aug>
    <au><snm>Vidrighin</snm><fnm>MD</fnm></au>
    <au><snm>Donati</snm><fnm>G</fnm></au>
    <au><snm>Genoni</snm><fnm>MG</fnm></au>
    <au><snm>Jin</snm><fnm>XM</fnm></au>
    <au><snm>Kolthammer</snm><fnm>WS</fnm></au>
    <au><snm>Kim</snm><fnm>M. S.</fnm></au>
    <au><snm>Datta</snm><fnm>A</fnm></au>
    <au><snm>Barbieri</snm><fnm>M</fnm></au>
    <au><snm>Walmsley</snm><fnm>IA</fnm></au>
  </aug>
  <source>Nat Commun</source>
  <publisher>Nature Publishing Group, a division of Macmillan Publishers
  Limited. All Rights Reserved.</publisher>
  <pubdate>2014</pubdate>
  <volume>5</volume>
  <url>http://dx.doi.org/10.1038/ncomms4532</url>
</bibl>

<bibl id="B9">
  <title><p>Real-time quantum feedback control of atomic
  spin-squeezing</p></title>
  <aug>
    <au><snm>Geremia</snm><fnm>J.</fnm></au>
    <au><snm>Stockton</snm><fnm>J.</fnm></au>
    <au><snm>Mabuchi</snm><fnm>H.</fnm></au>
  </aug>
  <source>Science</source>
  <pubdate>2004</pubdate>
  <issue>304</issue>
  <fpage>270</fpage>
  <lpage>273</lpage>
</bibl>

<bibl id="B10">
  <title><p>A hybrid on-chip optomechanical transducer for ultrasensitive force
  measurements</p></title>
  <aug>
    <au><snm>Gavartin</snm><fnm>E.</fnm></au>
    <au><snm>Verlot</snm><fnm>P.</fnm></au>
    <au><snm>Kippenberg</snm><fnm>T. J.</fnm></au>
  </aug>
  <source>Nat Nano</source>
  <publisher>Nature Publishing Group</publisher>
  <pubdate>2012</pubdate>
  <volume>7</volume>
  <issue>8</issue>
  <fpage>509</fpage>
  <lpage>-514</lpage>
  <url>http://dx.doi.org/10.1038/nnano.2012.97</url>
</bibl>

<bibl id="B11">
  <title><p>Minimum Requirements for Feedback Enhanced Force
  Sensing</p></title>
  <aug>
    <au><snm>Harris</snm><fnm>GI</fnm></au>
    <au><snm>McAuslan</snm><fnm>DL</fnm></au>
    <au><snm>Stace</snm><fnm>TM</fnm></au>
    <au><snm>Doherty</snm><fnm>AC</fnm></au>
    <au><snm>Bowen</snm><fnm>WP</fnm></au>
  </aug>
  <source>Phys. Rev. Lett.</source>
  <publisher>American Physical Society</publisher>
  <pubdate>2013</pubdate>
  <volume>111</volume>
  <fpage>103603</fpage>
  <url>http://link.aps.org/doi/10.1103/PhysRevLett.111.103603</url>
</bibl>

<bibl id="B12">
  <title><p>Ziv-Zakai Error Bounds for Quantum Parameter Estimation</p></title>
  <aug>
    <au><snm>Tsang</snm><fnm>M</fnm></au>
  </aug>
  <source>Phys. Rev. Lett.</source>
  <publisher>American Physical Society</publisher>
  <pubdate>2012</pubdate>
  <volume>108</volume>
  <fpage>230401</fpage>
  <url>http://link.aps.org/doi/10.1103/PhysRevLett.108.230401</url>
</bibl>

<bibl id="B13">
  <title><p>Biological measurement beyond the quantum limit</p></title>
  <aug>
    <au><snm>Taylor</snm><fnm>MA</fnm></au>
    <au><snm>Janousek</snm><fnm>J</fnm></au>
    <au><snm>Daria</snm><fnm>V</fnm></au>
    <au><snm>Knittel</snm><fnm>J</fnm></au>
    <au><snm>Hage</snm><fnm>B</fnm></au>
    <au><cnm>BachorHans A.</cnm></au>
    <au><snm>Bowen</snm><fnm>WP</fnm></au>
  </aug>
  <source>Nat Photon</source>
  <publisher>Nature Publishing Group</publisher>
  <pubdate>2013</pubdate>
  <volume>7</volume>
  <issue>3</issue>
  <fpage>229</fpage>
  <lpage>-233</lpage>
  <url>http://dx.doi.org/10.1038/nphoton.2012.346</url>
</bibl>

<bibl id="B14">
  <title><p>Adaptive homodyne measurement of optical phase</p></title>
  <aug>
    <au><snm>Armen</snm><fnm>M.A.</fnm></au>
    <au><snm>Au</snm><fnm>J.K.</fnm></au>
    <au><snm>Stockton</snm><fnm>J.K.</fnm></au>
    <au><snm>Doherty</snm><fnm>A.C.</fnm></au>
    <au><snm>Mabuchi</snm><fnm>H.</fnm></au>
  </aug>
  <source>Phys. Rev. Lett.</source>
  <pubdate>2002</pubdate>
  <volume>89</volume>
  <fpage>133602</fpage>
</bibl>

<bibl id="B15">
  <title><p>Adaptive Optical Phase Estimation Using Time-Symmetric Quantum
  Smoothing</p></title>
  <aug>
    <au><snm>Wheatley</snm><fnm>T.A.</fnm></au>
    <au><snm>Berry</snm><fnm>D.W.</fnm></au>
    <au><snm>Yonezawa</snm><fnm>H.</fnm></au>
    <au><snm>Nakane</snm><fnm>D.</fnm></au>
    <au><snm>Arao</snm><fnm>H.</fnm></au>
    <au><snm>Pope</snm><fnm>D.T.</fnm></au>
    <au><snm>Ralph</snm><fnm>T.C.</fnm></au>
    <au><snm>Wiseman</snm><fnm>H.M.</fnm></au>
    <au><snm>Furusawa</snm><fnm>A.</fnm></au>
    <au><snm>Huntington</snm><fnm>E.H.</fnm></au>
  </aug>
  <source>Phys. Rev. Lett.</source>
  <pubdate>2010</pubdate>
  <volume>104</volume>
  <issue>093601</issue>
</bibl>

<bibl id="B16">
  <title><p>Quantum-Enhanced Optical-Phase Tracking</p></title>
  <aug>
    <au><snm>Yonezawa</snm><fnm>H</fnm></au>
    <au><snm>Nakane</snm><fnm>D</fnm></au>
    <au><snm>Wheatley</snm><fnm>TA</fnm></au>
    <au><snm>Iwasawa</snm><fnm>K</fnm></au>
    <au><snm>Takeda</snm><fnm>S</fnm></au>
    <au><snm>Arao</snm><fnm>H</fnm></au>
    <au><snm>Ohki</snm><fnm>K</fnm></au>
    <au><snm>Tsumura</snm><fnm>K</fnm></au>
    <au><snm>Berry</snm><fnm>DW</fnm></au>
    <au><snm>Ralph</snm><fnm>TC</fnm></au>
    <au><snm>Wiseman</snm><fnm>HM</fnm></au>
    <au><snm>Huntington</snm><fnm>EH</fnm></au>
    <au><snm>Furusawa</snm><fnm>A</fnm></au>
  </aug>
  <source>Science</source>
  <pubdate>2012</pubdate>
  <volume>337</volume>
  <issue>6101</issue>
  <fpage>1514</fpage>
  <lpage>1517</lpage>
  <url>http://www.sciencemag.org/content/337/6101/1514.abstract</url>
</bibl>

<bibl id="B17">
  <title><p>Quantum-Limited Mirror-Motion Estimation</p></title>
  <aug>
    <au><snm>Iwasawa</snm><fnm>K</fnm></au>
    <au><snm>Makino</snm><fnm>K</fnm></au>
    <au><snm>Yonezawa</snm><fnm>H</fnm></au>
    <au><snm>Tsang</snm><fnm>M</fnm></au>
    <au><snm>Davidovic</snm><fnm>A</fnm></au>
    <au><snm>Huntington</snm><fnm>E</fnm></au>
    <au><snm>Furusawa</snm><fnm>A</fnm></au>
  </aug>
  <source>Phys. Rev. Lett.</source>
  <publisher>American Physical Society</publisher>
  <pubdate>2013</pubdate>
  <volume>111</volume>
  <fpage>163602</fpage>
  <url>http://link.aps.org/doi/10.1103/PhysRevLett.111.163602</url>
</bibl>

<bibl id="B18">
  <title><p>A Guide to Experiments in Quantum Optics</p></title>
  <aug>
    <au><snm>Bachor</snm><fnm>H A.</fnm></au>
    <au><snm>Ralph</snm><fnm>T. C.</fnm></au>
  </aug>
  <publisher>Weinheim: Wiley-VCH</publisher>
  <edition>2</edition>
  <pubdate>2004</pubdate>
</bibl>

<bibl id="B19">
  <title><p>Optimal State Estimation: Kalman, H Infinity, and Nonlinear
  Approaches</p></title>
  <aug>
    <au><snm>Simon</snm><fnm>D</fnm></au>
  </aug>
  <publisher>Hoboken: Wiley</publisher>
  <pubdate>2006</pubdate>
  <url>http://books.google.com.sg/books?id=urhgTdd8bNUC</url>
</bibl>

<bibl id="B20">
  <title><p>Linear Quadratic Control : An Introduction</p></title>
  <aug>
    <au><snm>Dorato</snm><fnm>P.</fnm></au>
    <au><snm>Abdallah</snm><fnm>C. T.</fnm></au>
    <au><snm>Cerone</snm><fnm>V.</fnm></au>
  </aug>
  <publisher>New York: MacMillan</publisher>
  <pubdate>1995</pubdate>
</bibl>

<bibl id="B21">
  <title><p>Lasers</p></title>
  <aug>
    <au><snm>Siegman</snm><fnm>A. E.</fnm></au>
  </aug>
  <publisher>University Science Books</publisher>
  <editor>A. Kelly</editor>
  <pubdate>1986</pubdate>
</bibl>

<bibl id="B22">
  <title><p>Laser Phase and Frequency Stabilization Using an Optical
  Resonator</p></title>
  <aug>
    <au><snm>Drever</snm><fnm>R. W. P.</fnm></au>
    <au><snm>Hall</snm><fnm>J. L.</fnm></au>
    <au><snm>Kowalski</snm><fnm>F. V.</fnm></au>
    <au><snm>Hough</snm><fnm>J.</fnm></au>
    <au><snm>Ford</snm><fnm>G. M.</fnm></au>
    <au><snm>Munley</snm><fnm>A. J.</fnm></au>
    <au><snm>Ward</snm><fnm>H.</fnm></au>
  </aug>
  <source>Appl. Phys. B</source>
  <pubdate>1983</pubdate>
  <volume>31</volume>
  <fpage>97</fpage>
  <lpage>105</lpage>
</bibl>

<bibl id="B23">
  <title><p>An introduction to {P}ound--{D}rever--{H}all laser frequency
  stabilization</p></title>
  <aug>
    <au><snm>Black</snm><fnm>E.D.</fnm></au>
  </aug>
  <source>Am. J. Phys.</source>
  <pubdate>2001</pubdate>
  <volume>69</volume>
  <issue>1</issue>
  <fpage>79</fpage>
  <lpage>87</lpage>
</bibl>

<bibl id="B24">
  <title><p>Linear Control Systems</p></title>
  <aug>
    <au><snm>Rohrs</snm><fnm>CE</fnm></au>
    <au><snm>Melsa</snm><fnm>JL</fnm></au>
    <au><snm>Schultz</snm><fnm>DG</fnm></au>
  </aug>
  <publisher>Singapore: McGraw-Hill</publisher>
  <edition>International edition</edition>
  <pubdate>1993</pubdate>
</bibl>

</refgrp>
} 
}


\ifthenelse{\boolean{publ}}{\end{multicols}}{}



%




%

\end{document}